\documentclass[preprint,english,showpacs,12pt,floatfix]{revtex4}
\usepackage{babel,amsmath,amssymb,dcolumn}
\usepackage{hyperref}
\usepackage{amsfonts}
\usepackage{amssymb}
\usepackage{graphicx}
\usepackage{latexsym}
\usepackage{dcolumn}
\usepackage{epsfig}
\usepackage{subfigure}
\usepackage{float}
\usepackage{color}

\begin{document}

\title{Phase transitions and regions of stability in Reissner-Nordstr\"om holographic superconductors.}

\author{E. Abdalla$^a$ }
\email{eabdalla@usp.br}

\author{C. E. Pellicer $^a$}
\email{carlosep@fma.if.usp.br}

\author{Jeferson de Oliveira $^a$}
\email{jeferson@fma.if.usp.br}

\affiliation{$^a$Instituto de F\'{\i}sica, Universidade de S\~ao
Paulo, CP 66318, 05315-970, S\~ao Paulo, Brazil}

\author{A. B. Pavan $^b$}

\email{alan@unifei.edu.br}

\affiliation{$^b$Instituto de Ci\^{e}ncias Exatas, Universidade Federal de Itajub\'{a},
Av. BPS 1303 Pinheirinho, 37500-903 Itajub\'{a}, MG, Brazil}

\begin{abstract}
The phase transition of Reissner-Nordstr\"om $AdS_{4}$ interacting with a massive charged scalar 
field has been further revisited. We found exactly one stable and one unstable
quasinormal mode regions for the scalar field. The two of them are separated by the
first marginally stable solution.
\end{abstract}

\maketitle

%%%%%%%%%%%%%%%%%%%%%%%%%%%%%%%%%

The AdS/CFT correspondence \cite{malda} has been used as a very important tool to probe physics
beyond the usual application in string theory, namely quark gluon plasma, fluid mechanics
as well as condensed matter models \cite{starinets, starinets2,gubser,gregorkanno}.
In the present discussion we analyse a model first considered in
\cite{gubser} and later by many authors \cite{gregorkanno,alanjef,binetal,maeda,siani,murataetal} related to
superconductors and corresponding phase transitions, generally referred to as holographic superconductors.

The AdS/CFT correspondence relates the gravitational fields in a bulk to properties of a field theory
at its border. Moreover, the weak coupling in one of the counterparts becomes the strong coupling
limit  of the other. In \cite{gubser,gubser1} a break of the $U(1)$ gauge symmetry has been achieved
by a charged scalar  interacting with a black hole in the bulk, a mechanism leading, eventually,
to a superconductor at the boundary --- what has been called a holographic superconductor.
Several similar mechanisms have been proposed in the literature. In particular, it has been
shown that several phase transitions may occur
in a given model \cite{gubser}. Such a
conclusion has been formulated upon computing the perturbations in the bulk and looking for
the time independent perturbations, which presumably divide sectors of growing and decaying modes.

Here we show that there are just two sectors, namely a superconducting and a normal phase, in view
of the fact that even beyond the marginally stable solution growing modes still exist.
Such a result comes about from the existence of further unstable modes, which we were able to
compute explicitly by the Quasi Normal Modes technique \cite{qnm}.

We consider the bulk action of a massive charged scalar field $\psi$ interacting with an Abelian
field $A_{\mu}$ generated by a charged  AdS black hole. The Lagrange density is that of Einstein
gravity with a cosmological constant, a charged scalar and a gauge field, that is,
\begin{equation}
\label{bulklagrangian}
16\pi G {\cal L} = R + \frac 6 {L^2} -\frac{F^2_{\mu\nu}}{4}-|\partial_{\mu}\psi-i q A_{\mu}
\psi|^2-m^2|\psi|^2\quad .
\end{equation}
The above Lagrangian has also been used in \cite{gubser} in the same context. The background spacetime
is a spherically symmetric Reissner-Nordstr\"om $AdS_{4}$, 
that is,
\begin{equation}
\label{metric}
ds^2 = -f(r)\ dt^2+\frac{1}{f(r)}\ dr^2+r^2\ d\Omega^2_{2}\quad,
\end{equation}
with
\begin{equation}
f(r)=1-\frac{2M}{r}+\frac{Q^{2}}{4r^{2}}+\frac{r^2}{L^{2}}\quad,
\end{equation}
where $Q$ and $M$ are respectively the electric charge and mass of the black hole. Besides,
$L$ refers to the AdS radius, which is related to the cosmological constant $\Lambda$ by
$L=\sqrt{-3/\Lambda}$. The Hawking temperature of this black hole is given by
\begin{equation}\label{temp}
T=\frac{1}{4\pi}\left[\frac{4r_{+}^{2}-Q^{2}}{4r_{+}^{3}}+\frac{3r_{+}}{L^{2}}\right]\quad.
\end{equation}
One identifies the parameters of the black hole with a critical temperature $T_{*}$
that characterize the phase transition. In \cite{superH}, this $T_{*}$ can be related to the
critical temperature $T_{c}$ of a holographic superconductor.

The solution for the gauge field $A_{\mu}$ reads
\begin{equation}\label{gauge}
A_{\mu}dx^{\mu}=\Phi(r) dt\quad,
\end{equation}
where the electric potential is written in terms of black hole charge $Q$ and the position of event
 horizon $r_{+}$ as
\begin{equation}
\Phi=\frac{Q}{r}-\frac{Q}{r_{+}}\quad.
\end{equation}

For our purposes, we assume that the scalar field $\psi$ is a small perturbation that does not
backreact. In this regime, we can use the quasinormal modes techniques in order to compute
the $\psi$-propagation. Its equation of motion results in
\begin{equation}
\label{eq1}
D_{\mu}D^{\mu}\psi=m^2\psi\quad,
\end{equation}
where $D_{\mu}=\nabla_{\mu}-iqA_{\mu}$. The parameters $m$ and $q$ are the mass and charge of the
scalar field, respectively. Expanding Eq.(\ref{eq1}) we find
\begin{eqnarray}
\label{eq2}
\Box \psi-2iqA_{\mu}g^{\mu\nu}\partial_{\nu}\psi-q^2 A_{\mu}A^{\mu}\psi-m^2\psi=0\quad .
\end{eqnarray}
The s-wave mode ($l=0$ mode in the angular momentum expansion) is easily obtained.
For our purposes
it is sufficient since higher angular momentum modes are connected with higher decay modes, and we
only need the first unstable mode to install unstability, as we are going to argue more generally.
We follow \cite{gubser} and set $r_{+}=Q=1$ and our parameters are $m$, $q$ and $L$.
We find
\begin{eqnarray}
\label{eq3}
-\frac{\partial^2\psi}{\partial t^2}+f^2\frac{\partial^2\psi}{\partial r^2}+\frac{\partial\psi}{\partial r}
\left[\frac{2f^2}{r}+f\frac{df}{dr}\right]+2iq\Phi\frac{\partial\psi}{\partial t}+q^2\Phi^2\psi-m^2f
\psi=0\quad .
\end{eqnarray}
We use the tortoise coordinate $dr_{*}=\frac{dr}{f}$ and the {\it{Ansatz}} $\psi=\frac{\Psi}{r}$ for
the field, resulting in the
equation
\begin{eqnarray}
\label{eq5}
-\frac{\partial^2\Psi}{\partial t^2}+\frac{\partial^2\Psi}{\partial r_{*}^2}+2iq\Phi
\frac{\partial\Psi}{\partial t}
- V(r) \Psi=0\quad ,
\end{eqnarray}
with
\begin{eqnarray}
\label{eqpotential}
V(r) = f(r) \left[ \frac{f^\prime}{r} + m^2 \right] - q^2 \Phi^2(r) \quad .
\end{eqnarray}

An analysis of Eq.(\ref{eq5}) shows that the procedure used in \cite{molina_wang} to discretize the spacetime
and integrate the equation of motion is not convenient because of the presence of the term
$2iq\Phi\frac{\partial\Psi}{\partial t}$. The method used in the mentioned work normally has numerical
error proportional to $\Delta^{4}$, where $\Delta$ is the step used, but the extra term has an error
proportional to $\Delta^{2}$, so we used the finite difference method, whose error is proportional to
$\Delta^{2}$ without the need of another coordinate change.

{\it Finite difference method }

We define
$\Psi(r_*,t) = \Psi(-j \Delta r_*,l \Delta t) = \Psi_{j,l}$, $V(r(r_*)) = V(-j \Delta r_*) = V_j$ and
$\Phi(r(r_*)) = \Phi(-j \Delta r_*) = \Phi_j$ to rewrite equation (\ref{eq5}) as
\begin{eqnarray}
&-& \frac{\left( \Psi_{j,l+1}-2\Psi_{j,l}+\Psi_{j,l-1}\right)}{\Delta t^2} + 2 i q \Phi_j
\frac{\Psi_{j,l+1}-\Psi_{j,l-1}}{2\Delta t} \nonumber \\
&+& \frac{\left( \Psi_{j+1,l}-2\Psi_{j,l}+\Psi_{j-1,l}\right)}{\Delta r_{*}^2} - V_j\Psi{j,l} +
O(\Delta t^2) + O(\Delta r_{*}^2) = 0 \quad .
\end{eqnarray}
With a Gaussian distribution with finite support as initial condition and Dirichlet conditions
at $r_*=0$, we derive the evolution of $\Psi$ by
\begin{eqnarray}\label{fdm}
\Psi_{j,l+1}=\left(1 - i q \Phi_j\Delta t\right)^{-1} \left[ - \left( 1 + i q
\Phi_j \Delta t \right) \Psi_{j,l-1} +
\frac{\Delta t^2}{\Delta r_{*}^2} \left( \Psi_{j+1,l} + \Psi_{j-1,l} \right) + \right.\nonumber\\
\nonumber\\
+\left. \left( 2 - 2 \frac{\Delta t^2}{\Delta r_{*}^2}- \Delta t^2 V_j \right) \Psi_{j,l} \right] \quad .
\end{eqnarray}

The Von Neumann stability conditions usually require that $\frac{\Delta t}{\Delta r_{*}}<1$. If the effective
potential Eq. (\ref{eqpotential}) is too large, the method is unstable even for small $\Delta t$. In a simple
wave equation $\ddot{x} + \omega^{2}x=0$, this requirement means that the step $\Delta t$ must be smaller
than $\omega^{-1}$. In our case, $V(r)$ is proportional to $\frac{r^{2}}{L^{4}}\left(2+m^{2}L^{2}\right)$ for
large $r$ (i.e. $r_{*}$ close to $0$), and for $m^{2}L^{2}=4$ we can see that even with ratios as low as
$\frac{\Delta t}{\Delta r_{*}}=0.7$ the method becomes  unstable, so we used $\frac{\Delta t}{\Delta r_{*}}=0.5$.
Therefore, in this work the stability conditions depend on the parameters $m$ and $L$.

%%%%%%%%%%%%%%%%%%%%%%%%%%%

{\it Horowitz-Hubeny method}

Another method to find quasinormal modes in asymptotically $AdS$ spacetimes was developed by 
Horowitz and Hubeny \cite{HH} and used in \cite{wanglinabd,konoplya}.

Rewriting the metric Eq.(\ref{metric}) and the gauge field Eq.(\ref{gauge}) in terms of
ingoing Eddington-Finkelstein coordinate, $v=t+r_{*}$, results in
\begin{equation}
\label{metricEdd}
ds^2=-f(r)\ dv^2+2\ dv\ dr +r^2\ d\Omega^2_{2}, \qquad A_{v}dv=\Phi(r) dv\quad.
\end{equation}
 In the new coordinates and after performing the separation of
variables $\psi(v,r,\theta,\phi)=\frac{Z(r)}{r}\ Y_{lm}(\theta,\phi)\ e^{-i\omega v}$
the wave equation Eq.(\ref{eq3}) reads
\begin{equation}
\label{eqEdd}
f\ \frac{d^2Z}{dr^2}+ \left[f'-2i\left(\omega+q\Phi\right)\right]\ \frac{dZ}{dr} -V(r)\ Z =0\quad,
\end{equation}
where the potential is
\begin{eqnarray}
\label{potEdd}
V(r)=\left(m^2+\frac{f'}{r}-\frac{2iq\Phi}{r}+\frac{\ell(\ell+1)}{r^2}\right)\quad.
\end{eqnarray}

Following the procedure outlined by Horowitz and Hubeny \cite{HH}, it is convenient to rewrite the
wave equation (\ref{eqEdd}) as
\begin{equation}
\label{eqHHx}
s(x)\frac{d^2Z}{dx^2}+\frac{t(x)}{(x-x_{+})}\ \frac{dZ}{dx}+\frac{u(x)}{(x-x_{+})^2}Z=0\quad,
\end{equation}
where we were be able to put the functions $s(x)$, $t(x)$ and $u(x)$ in a polynomial form
\begin{eqnarray}
\label{fHHx}
s(x)&=& x^2\ A_{0}\quad ,\\
t(x)&=& \left\{2(x-x_{+})x\ A_{0}+x\ A_{1}+2ix^2\left[\omega+q(x-x_{+})\right]\frac{}{}\right\}\quad ,\\
u(x)&=& (x-x_{+})\left\{A_{1}-m^2+2iq(x-x_{+}) x\frac{}{}\right\}\quad,
\end{eqnarray}
with
\begin{eqnarray}
\label{fHHx1}
A_{0}&=&\left[\frac{x^3}{4}-\frac{kx^2}{x_{+}}-\frac{x^2+x_{+}x+x_{+}^2}{L^2x_{+}^{3}}\right],\\
\nonumber\\
A_{1}&=&\left[-\frac{kx^3}{x_{+}}+\frac{(2x-x_{+})}{4}-\frac{(x^3+2x_{+}^{3})}{L^2x^3_{+}}\right].
\end{eqnarray}
where we set $l=0$ and $Q=1$.

Expanding the solution $Z(x)$ to the wave equation (\ref{eqHHx}) around the event horizon
$x_{+}=1/r_{+}$, we have
\begin{equation}
\label{HHexpan}
Z(x)=\sum_{n=0}^{\infty}a_{n}(\omega)(x-x_{+})^{n}.
\end{equation}
The key point is to compute de roots of the equation $Z(x=0)=0$ following from the boundary
condition at spatial infinity. Actually, we have to truncate the de sum at an intermediate $n=N$
and check that for the greater $n$ in which the root converge. We compute the zeros $\omega$
of $\sum_{n=0}^{\infty}a_{n}(\omega)(-x_{+})^{n}$  using the software {\it{Mathematica}} and
also the routine {\it{zroots}} \cite{nr}.

{\it Discussion and Results}

\begin{figure*}[ht!]
\begin{center}
\leavevmode
\begin{eqnarray}
\epsfxsize=4.8truecm\rotatebox{-90}
{\epsfbox{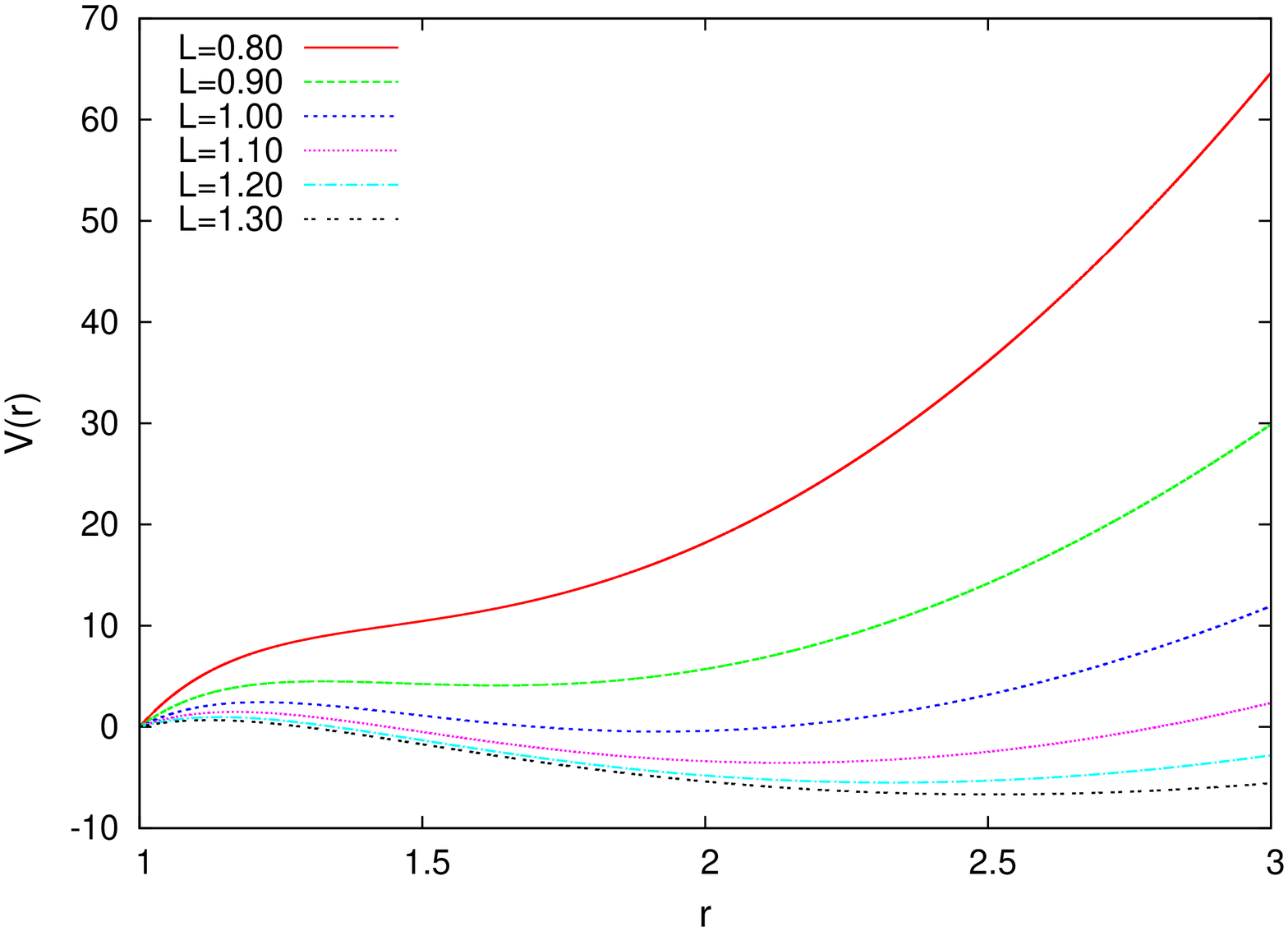}}\nonumber
\epsfxsize=4.8truecm\rotatebox{-90}
{\epsfbox{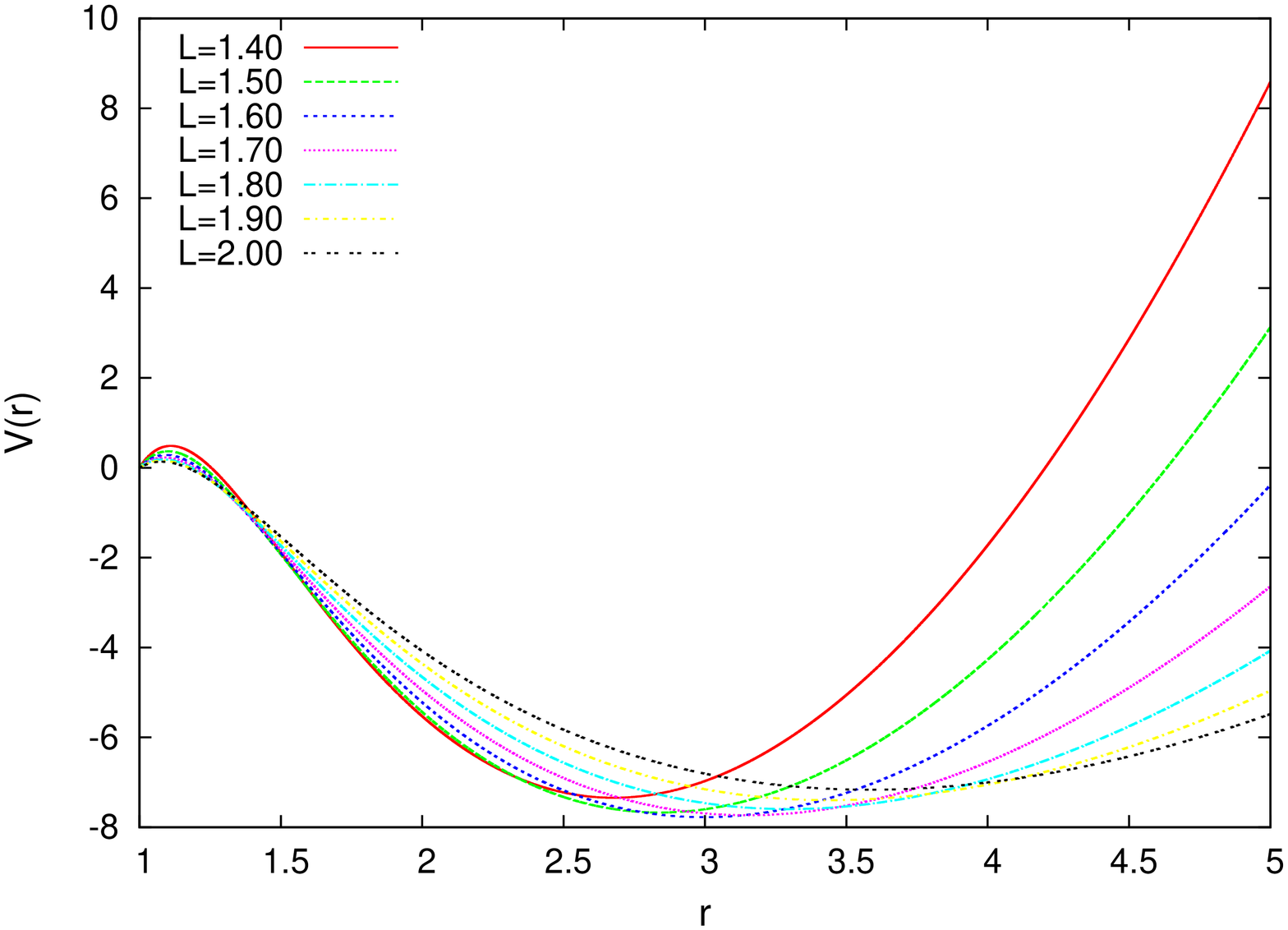}}\nonumber
\end{eqnarray}
\caption{{\small Behavior of effective potential for several values of $L$ (color online).}}
\label{figpotentials}
\end{center}
\end{figure*}

In Figure ({\ref{figpotentials}})
we plot  the effective potential (\ref{eqpotential}) due to the charged scalar field
propagation in the geometry (\ref{metric}) since the stable region,
passing through the transition point next to $L=1.28$ into the unstable region. We
clearly see that the potential defines a positive definite barrier near the event horizon
($r_{+}=1$) and becomes negative at intermediate values of radial coordinate
diverging at spatial infinite as expected for asymptotically AdS spacetimes. Since
we have an additional term in Eq.(\ref{eq5}), the result from \cite{HH} that a positive
definite potential implies stability is not necessarily valid. However, modes with
positive definite potentials are stable in this work. Besides, the first unstable mode
is presumably related to the first
value of $L$ (or highest value of $T$) whose potential admits a bound state with negative energy.

\begin{figure}[!ht]
\begin{center}
\epsfig{file=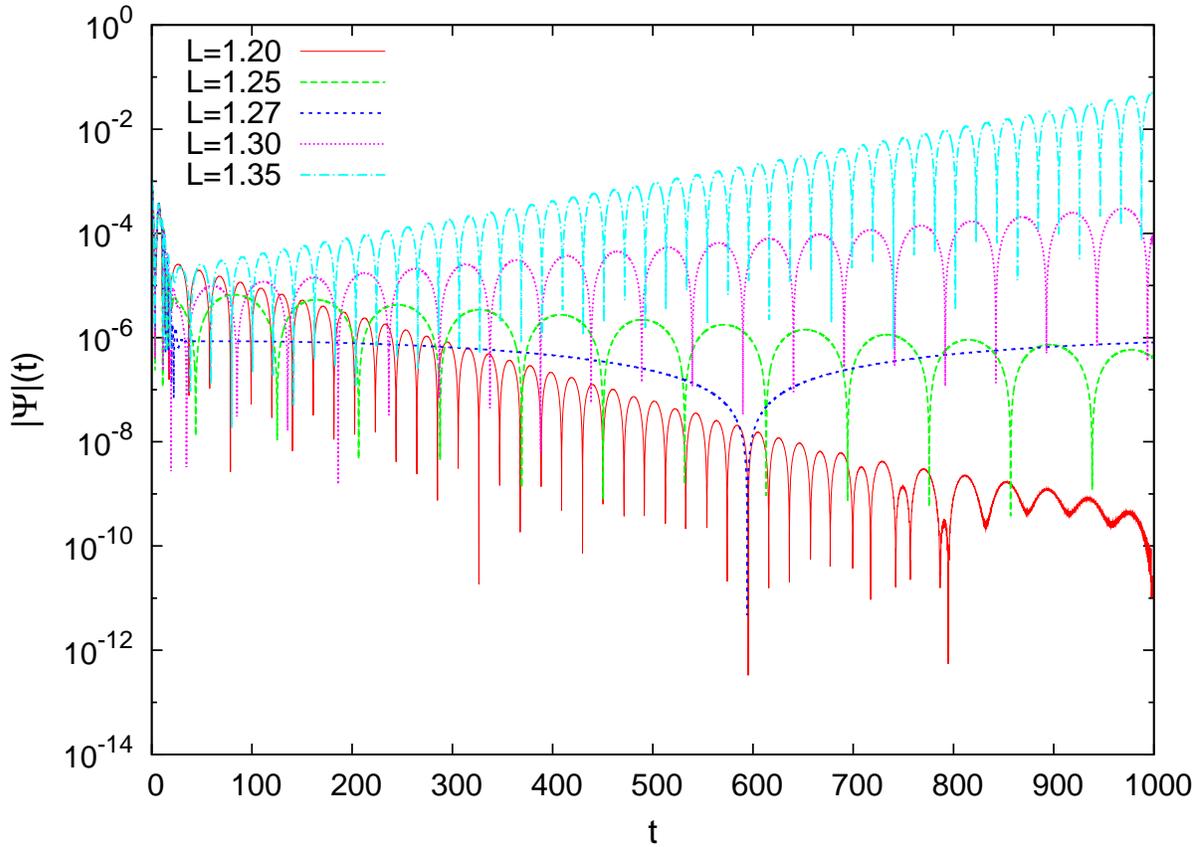, angle=-90, width=1.0 \linewidth, clip=}
\end{center}
\caption{{\small Quasinormal modes versus $t$ (color online).}}
\label{figtrans}
\end{figure}

Figure (\ref{figtrans}) shows five different behaviors of $\psi(r_*,t)$ given by Eq.(\ref{fdm})
at $r_*=-5$ for $m^2L^2=4$ and $qL=10$. The results show an interesting phenomena. At a value next
to $L=1.28$ we have a phase transition. This is the same transition obtained by
Gubser. Our results agree with those, that is, beyond this value of $L$ the black hole become unstable.

\begin{figure*}[ht!]
\begin{center}
\leavevmode
\begin{eqnarray}
\epsfxsize= 5.0truecm\rotatebox{-90}
{\epsfbox{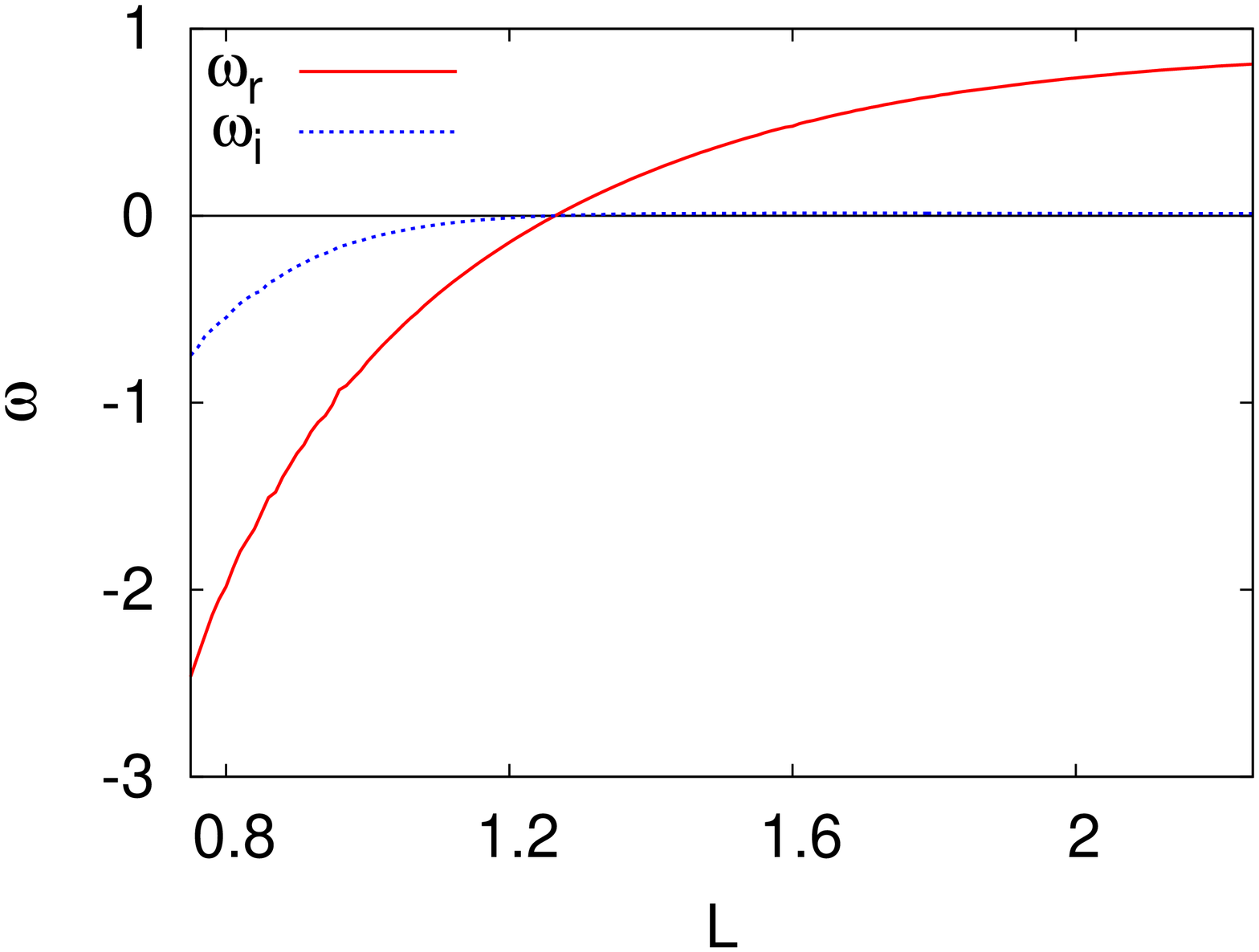}}\nonumber
\epsfxsize= 5.0truecm\rotatebox{-90}
{\epsfbox{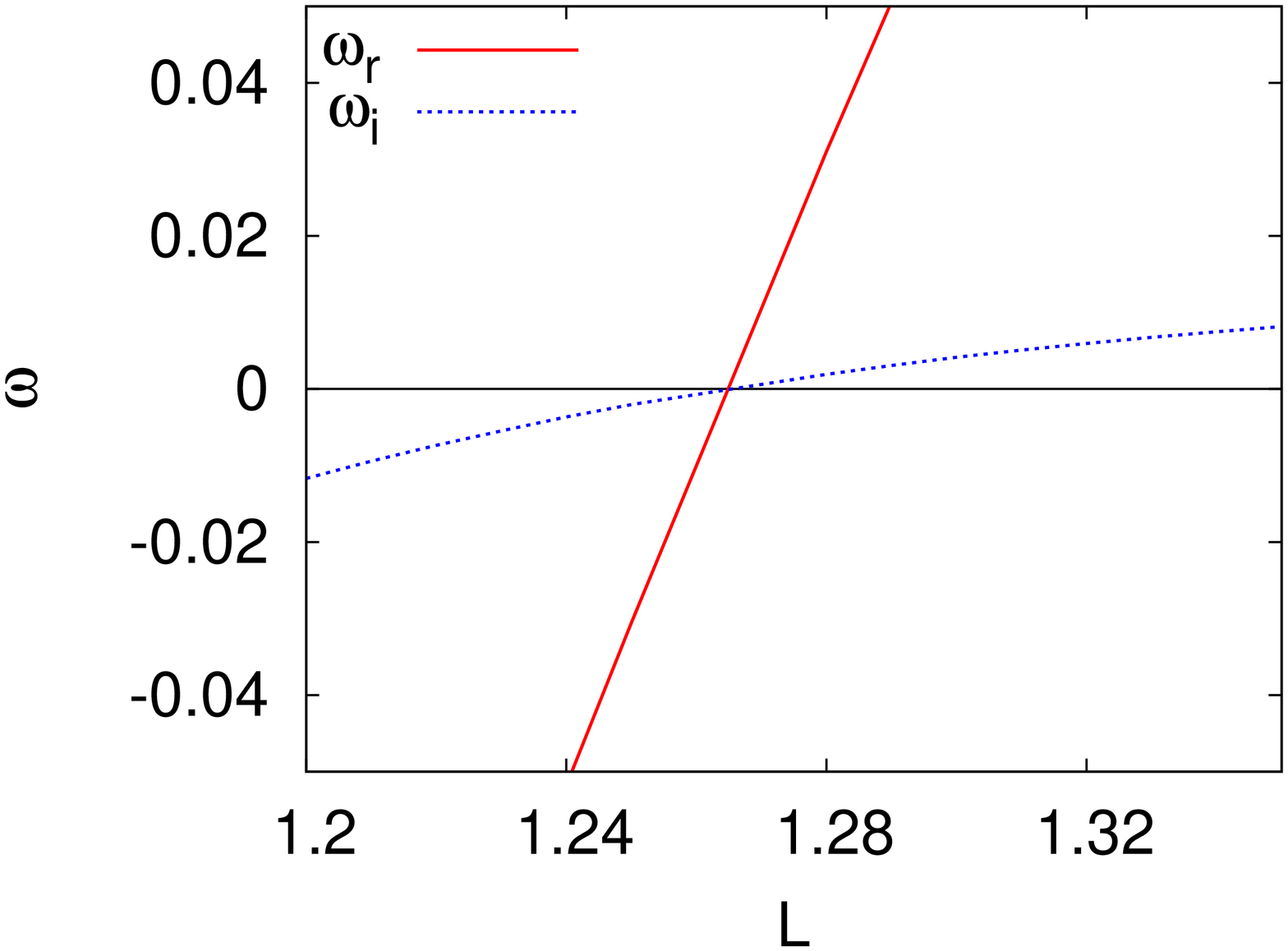}}\nonumber
\end{eqnarray}
\caption{{\small First mode frequencies (left) and near the transition (right). (color online)}}
\label{figfreq1L}
\end{center}
\end{figure*}

\begin{figure*}[ht!]
\begin{center}
\leavevmode
\begin{eqnarray}
\epsfxsize= 5.0truecm\rotatebox{-90}
{\epsfbox{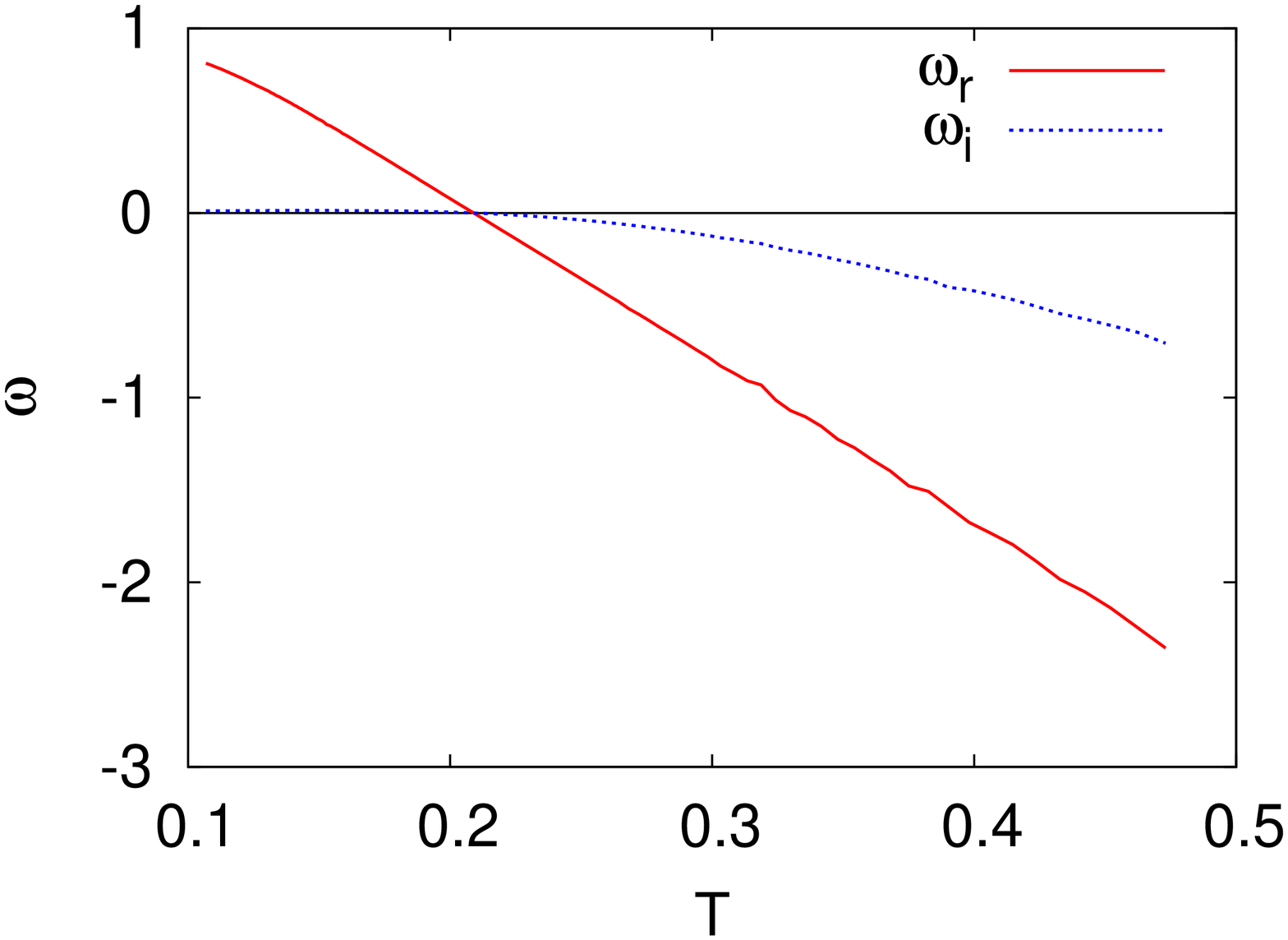}}\nonumber
\epsfxsize= 5.0truecm\rotatebox{-90}
{\epsfbox{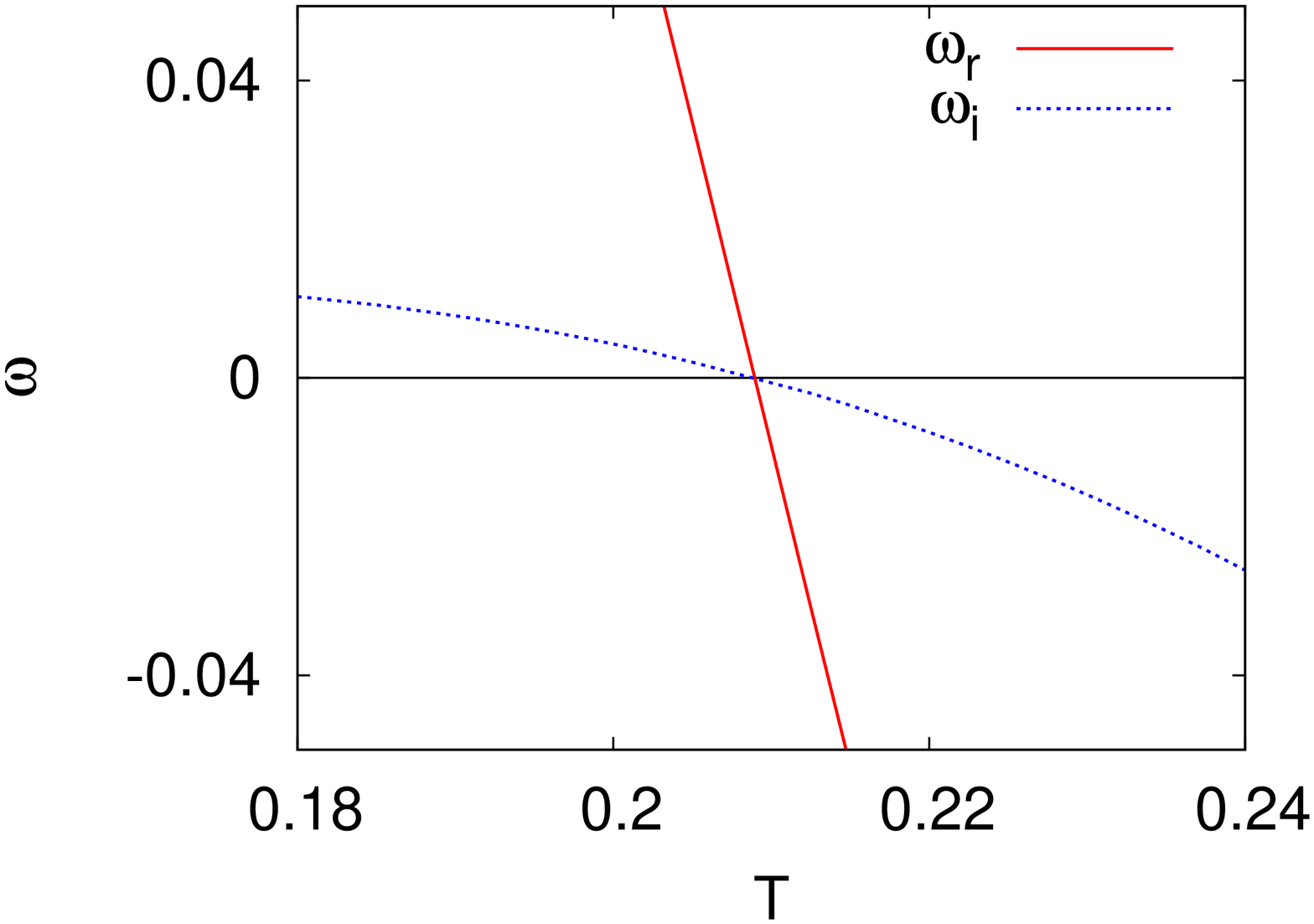}}\nonumber
\end{eqnarray}
\caption{{\small First mode frequencies in funcion of $T$ (left) and near the transition (right). (color online)}}
\label{figfreqT01}
\end{center}
\end{figure*}

Fitting $\Psi=A\exp(\omega_i t)\cos(\omega_r t + \delta)$, we get the quasinormal frequencies.
We cannot tell the sign of $\omega_r$ using this fitting, but if we
assume that $\omega_r$ depends smoothly on $L$, a sign change would appear as a discontinuity
in the first derivative, as our data appears. Choosing the sign of $\omega_r$ as the same
of $\omega_i$, we obtain the figure (\ref{figfreq1L}), which shows that both $\omega_r$ and
$\omega_i$ change sign at the same value of $L$.
In the same figure we adjust the scale so that the sign change is clearly seen.
We are also interested in the dependence of these frequencies with the temperature, shown in
figure (\ref{figfreqT01}), in this figure we also adjust the scale to better see the transition.
We noticed that the transition occurs at a value of $L$ smaller than expected in \cite{gubser}.
Since the numerical method used to solve second order partial differential equations has errors
larger than methods commonly used to solve second order ordinary differential equations, we believe
that this difference is due to numerical errors.

\begin{figure*}[ht!]
\begin{center}
\leavevmode
\begin{eqnarray}
\epsfxsize= 5.9truecm\rotatebox{-90}
{\epsfbox{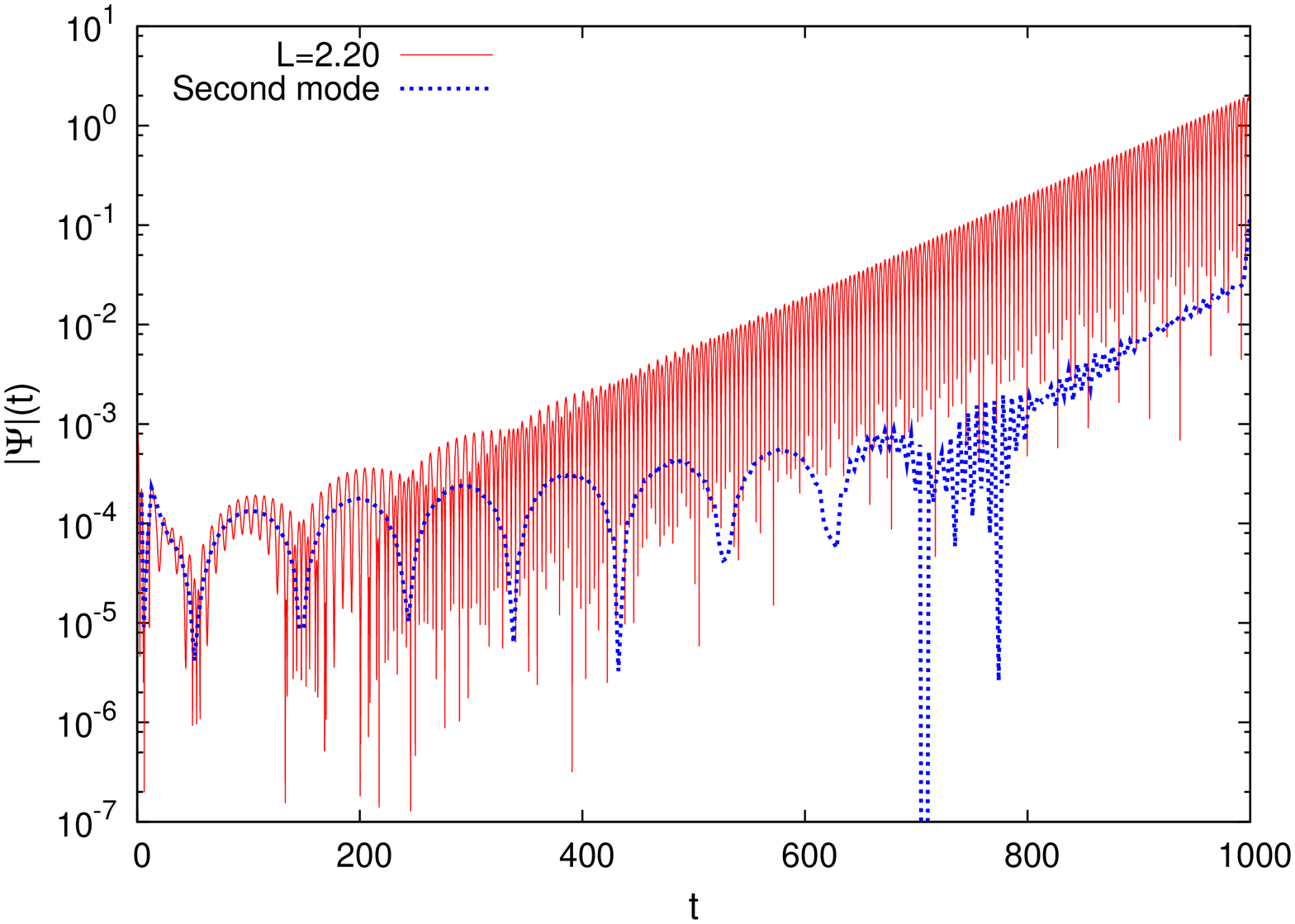}}\nonumber
\epsfxsize= 5.9truecm\rotatebox{-90}
{\epsfbox{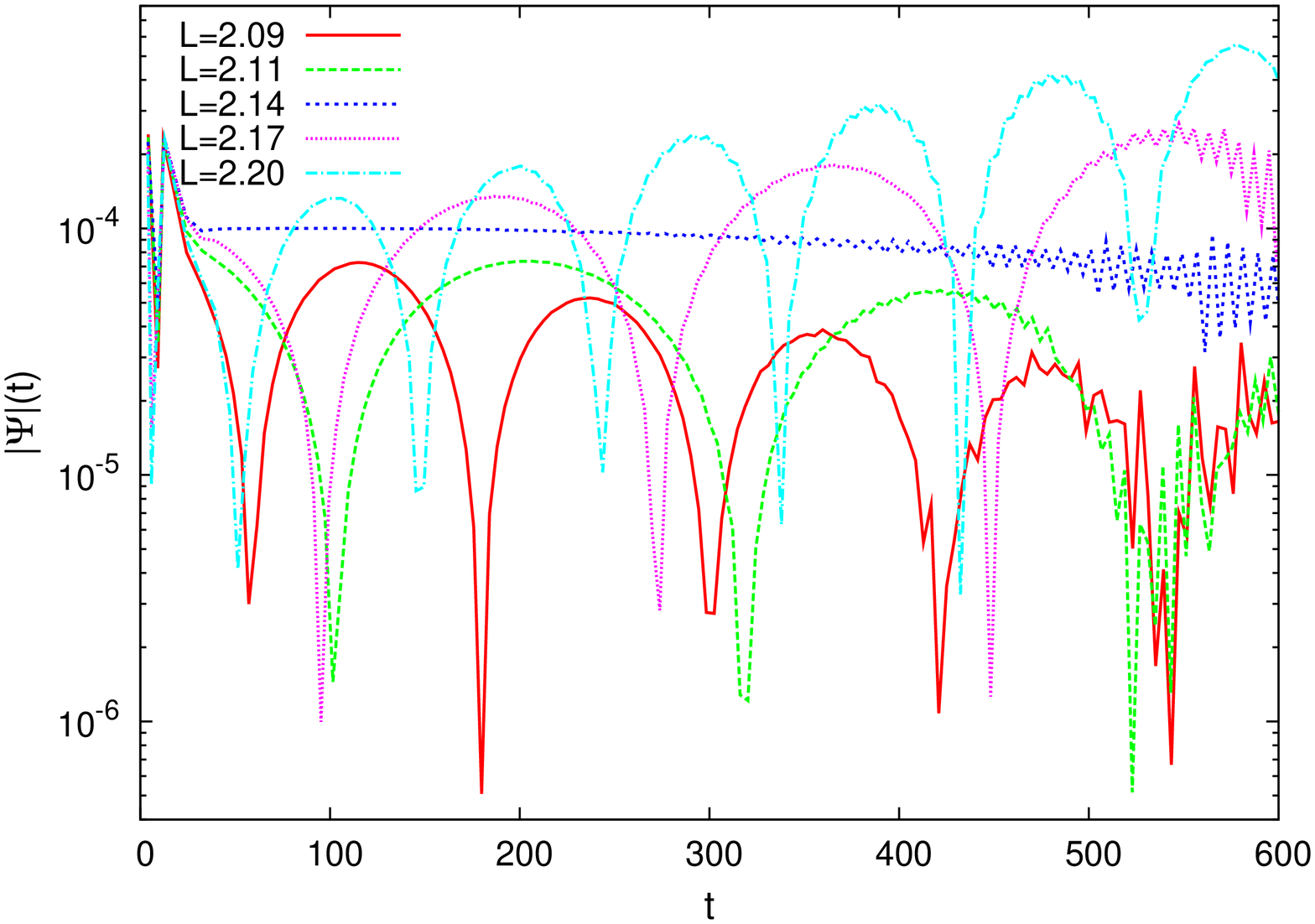}}\nonumber
\end{eqnarray}
\caption{\small Evolution of $\Psi(t)$ with isolated second mode (left) and second mode behaviours for several $L$ (right). (color online)}
\label{figmode2}
\end{center}
\end{figure*}

There are further secondary quasinormal modes. We found the next one which is presented in figure
(\ref{figmode2}). Indeed, the secondary mode shows a zero frequency at $L$ next to the second
marginally stable mode found in \cite{gubser}. However, the first mode remains, and it
is unstable. This means that the model remains unstable and superconductivity
still holds beyond that second marginally stable mode.

\begin{figure*}[ht!]
\begin{center}
\leavevmode
\begin{eqnarray}
\epsfxsize= 5.0truecm\rotatebox{-90}
{\epsfbox{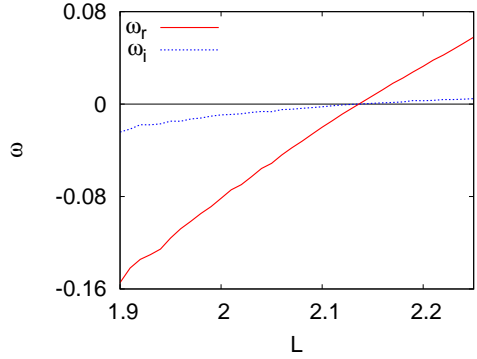}}\nonumber
\end{eqnarray}
\caption{{\small Second mode frequencies (color online).}}
\label{figfreq2}
\end{center}
\end{figure*}

The frequencies of the second mode displays a behaviour similar to the first mode. Again, choosing
the sign of $\omega_r$ as the same of $\omega_i$, we see that both frequencies change sign at
the same point. These frequencies are shown in Figure (\ref{figfreq2}).

\begin{figure*}[ht!]
\begin{center}
\leavevmode
\begin{eqnarray}
\epsfxsize= 5.0truecm\rotatebox{-90}
{\epsfbox{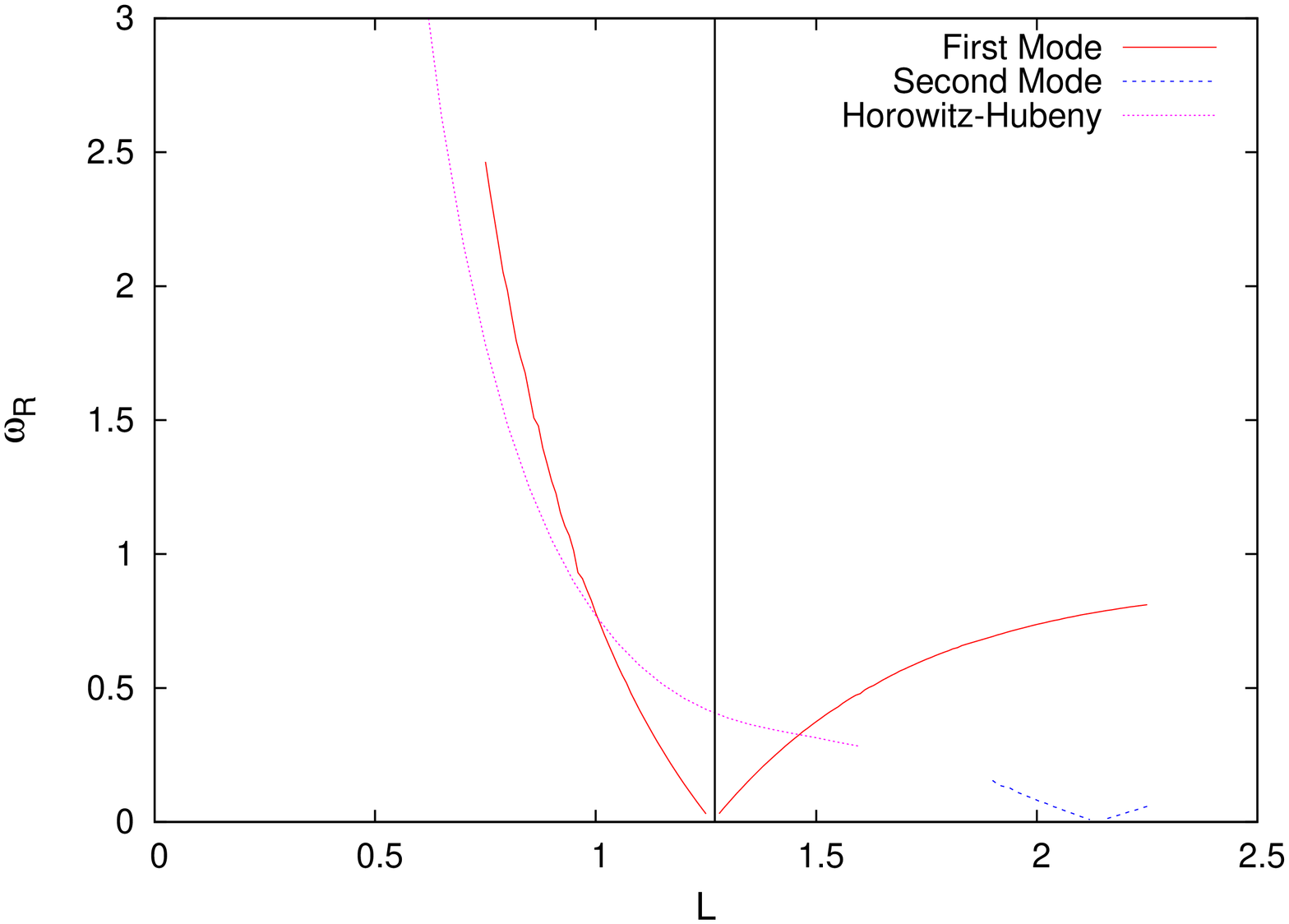}}\nonumber
\epsfxsize= 5.0truecm\rotatebox{-90}
{\epsfbox{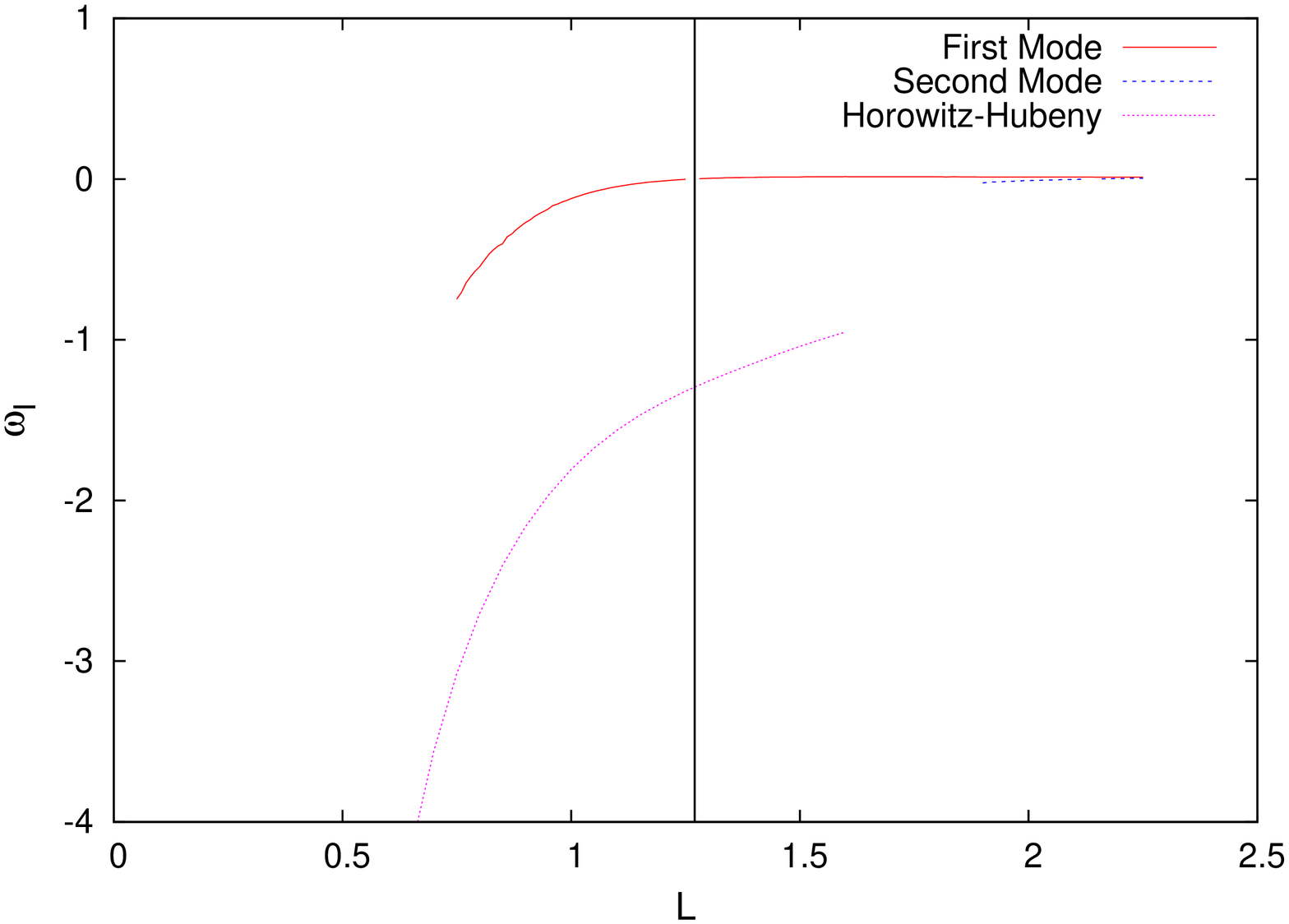}}\nonumber
\end{eqnarray}
\caption{{\small Comparison between real frequencies (left) and imaginary frequencies (right). (color online)}}
\label{figcomp}
\end{center}
\end{figure*}

In Figure (\ref{figcomp}) we have the comparison between the results obtained using
two different numerical approaches: the finite difference method Eq.(\ref{fdm}) and Horowitz-Hubeny method
Eq.(\ref{HHexpan}). The vertical line separates the stable and unstable regions.

In the stable region, the behavior of the
real part of the quasinormal frequencies as $L$ increases is quite similar in both approaches.
However, after the phase transition, the Horowitz-Hubeny method gives a rather different result.
A possible explanation for this behavior could be due to  the fact that in the neighborhood of the phase
 transition point $L_{*}$, the real part of the frequencies computed through this method may be the
frequencies of the second mode.
Indeed, zero frequencies do not show up in the Horowitz-Hubeny method due to the its limitations
as already discussed in \cite{HH}.
We also found some frequencies with positive imaginary part using the Horowitz-Hubeny method.
However, it does not seem to converge as we increase the truncation order $N$. Therefore, we cannot
rely on the results in this region.

According to \cite{acton}, polynomials with large terms (e.g. $\sim 10^{18}$) can be pathologic. A
vanishingly small change in a very small term can drastically change the roots in such cases.
 In the Horowitz-Hubeny method, the larger the truncation order $N$, the larger the terms of the
polynomial are. Thus, for sufficiently
large values of $N$, even small numerical errors from polynomial calculation can lead to a completely
different root. That explains why the roots stop converging after a certain $N$. In this work, the first
derivative term in (\ref{eq5}) implies in larger terms in the polynomial and the algorithm
stops converging for $N$ smaller than what is expected for systems with $q=0$. The approximate
positions of the zeros as well as stability are confirmed as well.

Therefore, we conclude that there are exactly two phases in the model, a stable phase for small
values of $L$ (normal phase) and a unstable superconducting phase for $L > 1.28$, or for
temperatures $T < 0.21$. We expect such results and phase structure to remain true in other
holographic models (as e.g. with different topologies).

{\it{Acknowledgments}}

This work has been supported by FAPESP ({\it{Funda\c c\~ao de Amparo \`a Pesquisa do Estado de S\~ao Paulo}}) 
and CNPq ({\it{Conselho Nacional de Desenvolvimento Cient\'ifico e Tecnol\'ogico}}), Brazil.

\newpage
%%%%%%%%%%%%%%%%%%%%%%%%%%%%%%%%%%%%%%%%%%%%%%%%%%%%%%%%%%%%%%%%%%%%%%%%%%%%%%

\end{document}